\documentclass[prl,aps,floatfix,twocolumn,showpacs,preprintnumbers,
superscriptaddress]{revtex4}
\usepackage{graphicx}
\usepackage{dcolumn}
\usepackage{bm}
\usepackage{amssymb}
\usepackage{pifont}
\usepackage{amsmath}
\usepackage{times}
\usepackage[nooneline]{subfigure}
\usepackage{color}

\begin{document}

\title{From brittle to ductile fracture in disordered materials}
\author{Clara B.\ Picallo} 
\affiliation{Instituto de F\'{\i}sica de Cantabria (IFCA), CSIC--UC, E-39005
Santander, Spain}
\affiliation{Departamento de F{\'\i}sica Moderna, Universidad de Cantabria,
Avda. Los Castros, E-39005 Santander, Spain}
\author{Juan M.\ L{\'o}pez} 
\affiliation{Instituto de F\'{\i}sica de Cantabria (IFCA), CSIC--UC, E-39005
Santander, Spain}
\author{Stefano Zapperi} 
\affiliation{CNR-IENI, Via R. Cozzi 53, 20125 Milano, Italy }
\affiliation{ISI Foundation, Viale S. Severo 65, 10133 Torino, Italy}

\author{Mikko J.\ Alava} 
\affiliation{Aalto University, Department of Applied Physics, PO. Box 14100,
00076 Aalto, Finland}

\date{\today}

\begin{abstract}
We introduce a lattice model able to describe damage and yielding in
heterogeneous materials ranging from brittle to ductile ones. Ductile fracture
surfaces, obtained when the system breaks once the strain is completely
localized, are shown to correspond to minimum energy surfaces. The
similarity of the resulting fracture paths to the limits of brittle fracture or
minimum energy surfaces is quantified. The model exhibits a
smooth transition from brittleness to ductility, also present
in how much plastic deformation is accumulated prior to fracture. The
dynamics of yielding exhibits avalanches with a power-law distribution.

\end{abstract}

\pacs{62.20.F-, 62.20.M-, 05.40.-a, 61.43.-j}

\maketitle

Understanding and characterizing the complexity in material failure
is of great interest in both basic physics research and materials
science applications. Brittle materials fail in the elastic regime,
while ductile materials can locally accumulate plastic deformation
prior to fracture and they are often able to withstand higher
stresses before reaching fracture~\cite{greer95}. Numerical models
of fracture often involve molecular dynamics
simulations~\cite{shukla,buehler}. To go beyond these,
brittle fracture with its simpler rheology has been extensively
analyzed at the mesoscopic scale by both scalar and tensorial
lattice models~\cite{alava06}. This provides theoretical tools for
understanding the intermittency observed in fracture-- avalanches--
and the question why crack surfaces roughen, to name some key
problems~\cite{alava06}. Other mesoscale models have been developed
for plastic deformation and its accumulation in amorphous and
disordered materials~\cite{pp92,sor93a,sor93b,roux02,sor99}. In
plastic deformation one sees, in analogy to brittle fracture,
general scaling laws like acoustic
emission~\cite{miguel01,richeton06,wang09} and strain avalanche
distributions~\cite{dimiduk06,schwer07,zaiser08,sun10}. A relevant
phenomenon is the spatial localization of strain, sometimes
noticeable as shear bands that can then induce final
failure~\cite{li05,li07,yao08,schall07}. One important issue in plasticity
has been ``optimization'', as the yield surface may minimize the sum
of certain local random quenched variables~\cite{pp92,picallo09}.

In this Letter we study a scalar lattice model for fracture and
plastic deformation in elasto-plastic heterogeneous materials.
Depending on the material properties the model is able to exhibit
fracture in situations that range from brittle to ductile behavior.
In analogy to dislocation dynamics and other scenarios of plastic
deformation, for ductile systems the  dynamics of yielding is
characterized by strain
bursts~\cite{miguel01,richeton06,wang09,dimiduk06,schwer07,zaiser08}.
Finally, plastic deformation localizes into shear bands, until a
crack develops. The model is a generalization of the well-known
random fuse model (RFM)~\cite{arcangelis85} for brittle fracture.
Our ductile random fuse model (DRFM) is able to accumulate plastic
deformation before complete failure. Depending on a model parameter,
the model can interpolate between brittle failure or perfect
plasticity depending on how ductile the system is, or how much it
can yield. Due to the two intertwined dynamics of fracture and
yielding the DRFM presents a very rich behavior, which we explore
below.

\paragraph{The model.-}
The RFM represents a scalar lattice electrical analog of the elasticity problem
where the local stress ($\sigma_i$), strain ($\epsilon_i$), and local elastic
modulus ($E_i$) are mapped to the current ($I_i$), potential drop ($V_i$), and
local conductance ($g_i$), respectively, of a network of fuses subject to an
external voltage. Bus bar boundary conditions are imposed at the top and bottom
of the system and periodic boundary conditions in the transverse directions.
Each individual fuse $i$ behaves ohmic (elastic), $I_i = g_i V_i$ (equivalently,
$\sigma_i = E_i \epsilon_i  $), up to a local threshold current $T_i$, which is
a uniformly distributed quenched random variable.

\begin{figure}
\centerline{\includegraphics[width=75mm,type=eps,ext=.eps,read=.eps]{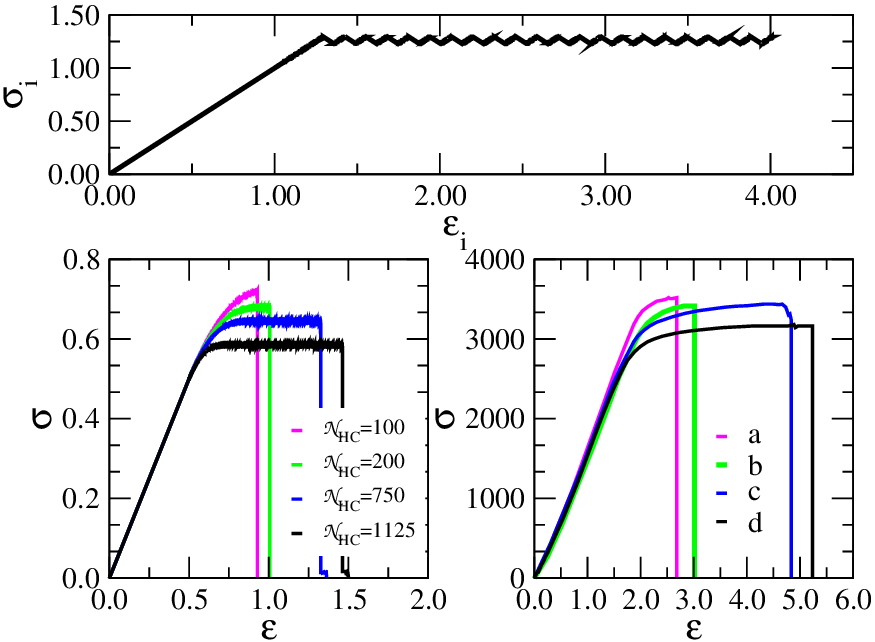}}
\caption{(Color online) DRFM in a diamond lattice of size $L=75$ with disordered
thresholds uniformly distributed in the interval $\left[0.5,1.5\right]$. Top
panel shows a typical stress-strain response curve for a single element of the
lattice for $\beta=0.1$. Left bottom panel shows global stress-strain curves for
different typical disorder realizations with an increasing number of healing
cycles and a decreasing median of the threshold distribution. For comparison,
right bottom panel shows experimental data for amorphous steel alloys with
increasing ductility from a to d (from Ref.~\cite{gu08}).
}
\label{fig1}
\end{figure}

The DRFM that we introduce here is based on the scalar tectonic model of Cowie
{\it et al.}~\cite{sor93b}, but it could readily be extended to tensorial models
with more degrees of freedom such as beam models~\cite{alava06}. Whenever a fuse
reaches its threshold $T_i$, a permanent deformation is imposed to the element
and it becomes elastic again. This defines a ``healing cycle'' of the individual
element. In the electrical analogue this healing is done by imposing a voltage
source ({\it i.e.} an electric battery) in series with the fuse to generate an
opposite current through it so that elastic deformation is relaxed below
threshold while plastic deformation accumulates in the element. The magnitude of
the local voltage source used to heal a fuse is linearly related to its
threshold, $\Delta_{i} = \beta \, T_i/g_{i}$ where $\beta$ is a parameter that
controls how much local deformation is allowed to be accumulated at each healing
cycle. We repeat this healing cycle a fixed number of times ${\cal N}_{HC}$ for
each fuse going above threshold (see top panel of Fig.~\ref{fig1}) until it
definitely breaks. The number of healing cycles ${\cal N}_{HC}$ that each
individual element can go through before failure is fixed for all the fuses and
parametrizes in a simple and convenient manner the yielding characteristics
(ductility) of the material. In our simulations we used $\beta=0.1$ and the
local elastic modulus or conductivity $g_i=1$, unless stated otherwise. Note
that in the double limit ${\cal N}_{HC} \to \infty$ and $\beta \to 0$, one would
obtain a elastic-perfectly plastic response. At each step of the simulation both
the fuse closest to threshold and the external voltage required to reach it can
be exactly calculated. Therefore, the external voltage is increased exactly up
to the point where the next fuse in the network reaches its threshold. The
current redistribution after the healing of a fuse can cause other fuses to also
reach their thresholds. Therefore, avalanches of plastic events are observed
similar to the strain bursts observed in
experiments~\cite{dimiduk06,ng09,wang09}.

Eventually fracture occurs as follows. At any given time in the simulation
different sites of the network have gone through a different number of healing
cycles, reflecting the spatially varying distribution of strain in the system. A
fuse is forced to burn (break) after having gone through a fixed number of
healing cycles ${\cal N}_{HC}$, so that this quantity parametrizes the
capability of the system to sustain local deformation. The fuse then
irreversibly becomes an insulator and all the voltage sources that were imposed
as plastic deformation are removed. When a continuous path of insulating bonds
is formed, the system is disconnected and fails completely.

\paragraph{Brittle versus ductile fracture surfaces.-} We have studied the DRFM
in two dimensions in a system of size $L_x \times L_y$. Bus bars at $y=0$ and
$y=L_y+1$ impose an external voltage $V$ across the system.
Figure 1 (top panel) shows the stress-strain curve for an
individual element of the network in the DRFM. Each stress drop corresponds to a
healing cycle of the element and the amplitude of the zig-zag is proportional to
$\beta$. Global stress in this model is defined as
$\sigma \equiv I/L_x$ and the strain is given by $\epsilon \equiv V/L_y$, where
$I$ is the total current, $V$ is the voltage drop, and $L_y=L_x/2+1=L+1$ for a
system of size L in a ``diamond'' ($45^o$ tilted square) lattice.
Figure~\ref{fig1} (left bottom panel) shows the global stress-strain curve for
$\beta=0.1$ as the number of healing cycles ${\cal N}_{HC}$ before breakdown is
increased and the median of the threshold distribution is progressively
reduced. Also, in Fig.~\ref{fig1} (right bottom panel) we plot stress-strain
curves obtained by Gu {\it et~al.}~\cite{gu08} in recent experiments with
amorphous steel alloys of of Fe-Cr-Mo-P-C-B with different ductility produced by
changing the metal-metalloid composition. The comparison illustrates how the
index ${\cal N}_{HC}$ can parametrize in a simple manner the ductility of the
experimental samples. Our results are also in excellent agreement with very
recent experiments by Sun {\it et~al.}~\cite{sun10} on ductile metallic glasses
showing the cycles of sudden stress drop followed by elastic reloading
associated with shear-band motions. 

\begin{figure}
\centerline{\includegraphics[width=75mm,type=eps,ext=.eps,read=.eps]{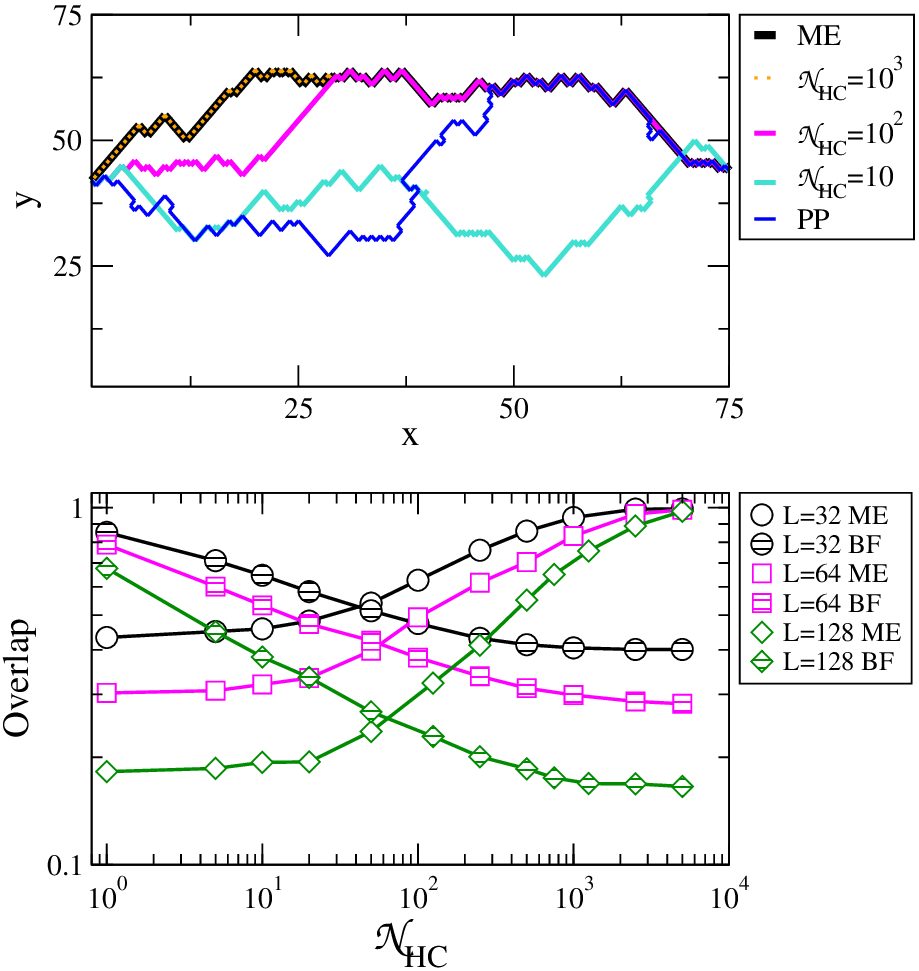}}
\caption{(Color online) Top: ME, PP and DRFM fracture surfaces for the {\em
same} disorder realization in a $L=75$ system. One can see that the final
fracture surface gets closer to ME as the system becomes more ductile. Bottom:
Average over $10^3$ disorder realizations of the total spatial overlap of the
DRFM final crack with both the corresponding ME and BF (${\cal N}_{HC} = 0$)
surfaces for $\beta=0.1$.
}
\label{fig2}
\end{figure}

Figure~\ref{fig2} (top panels) shows the resulting fracture paths in
the DRFM for the same conditions and disorder configuration as in
the simulations shown in Fig.~\ref{fig1} (bottom left panel). One
immediately notices that the final fracture surface configuration
depends on the number of healing cycles ${\cal N}_{HC}$ and
therefore on the accumulated plastic strain. These surfaces are to
be compared with the ones emerging from the perfect plasticity limit
(${\cal N}_{HC} \rightarrow \infty$)~\cite{picallo09}. As a
reference, we plot the minimum energy (ME) surface~\cite{mid95} and
the perfectly plastic (PP) path found with the algorithm of Roux and
Hansen~\cite{pp92,picallo09} for exactly the same disorder
configuration. These two surfaces are known to minimize the sum of
the local yield stresses and the stress flowing through the surface,
respectively~\cite{picallo09}. It becomes apparent that the deeper
the system is allowed into the plastic steady-state (${\cal N}_{HC}
\gg 1$), the closer the resulting fracture surface is to the ME path
for the same disorder configuration. We define the overlap between
two given paths $\{x_i\}$ and $\{y_i\}$ as $(1/Z) \sum_{i}
\delta(x_i-y_i)$, where $Z$ is a normalization constant so that the
overlap becomes unity for two identical surfaces. One can see in
Fig.~\ref{fig2} (top panel) that the fracture surface tends to
overlap with the ME surface as the ductility ${\cal N}_{HC}$
increases. Also, in Fig.~\ref{fig2} (bottom panel) we plot the
overlap between the plastic fracture surface obtained with the
directed ME surface for the same disorder realization. For the sake
of comparison, we also compute the overlap with  the corresponding
purely brittle fracture surface~\cite{arcangelis85} for the same
disorder realization ({\it i.e.}, setting ${\cal N}_{HC}=0$). The PP
surface seems to be irrelevant for this problem and, consequently,
the overlap is negligible for all ${\cal N}_{HC}$ (not shown).

Figure~\ref{fig2} demonstrates that the system is quasi-brittle for low values
of ${\cal N}_{HC}$ where the fracture surface largely overlaps with the purely
brittle fracture surface. In contrast, if the material is allowed to accumulate
locally more strain ({\it i.e.} for larger ${\cal N}_{HC}$) then the overlap
with the brittle fracture rapidly decreases, while the path becomes
progressively closer to the directed ME path. This is in agreement with the
results reported in more complex models~\cite{sor93a}. Moreover, the finite-$L$
behavior of the overlap in Fig.~\ref{fig2}, which decays as $\sim \log (1/L)$
with system size in both ${\cal N}_{HC} \to \infty$ (ME) and ${\cal N}_{HC} = 1$
(quasi-brittle) limits.
\begin{figure}
\centerline{\includegraphics[width=75mm,type=eps,ext=.eps,read=.eps]{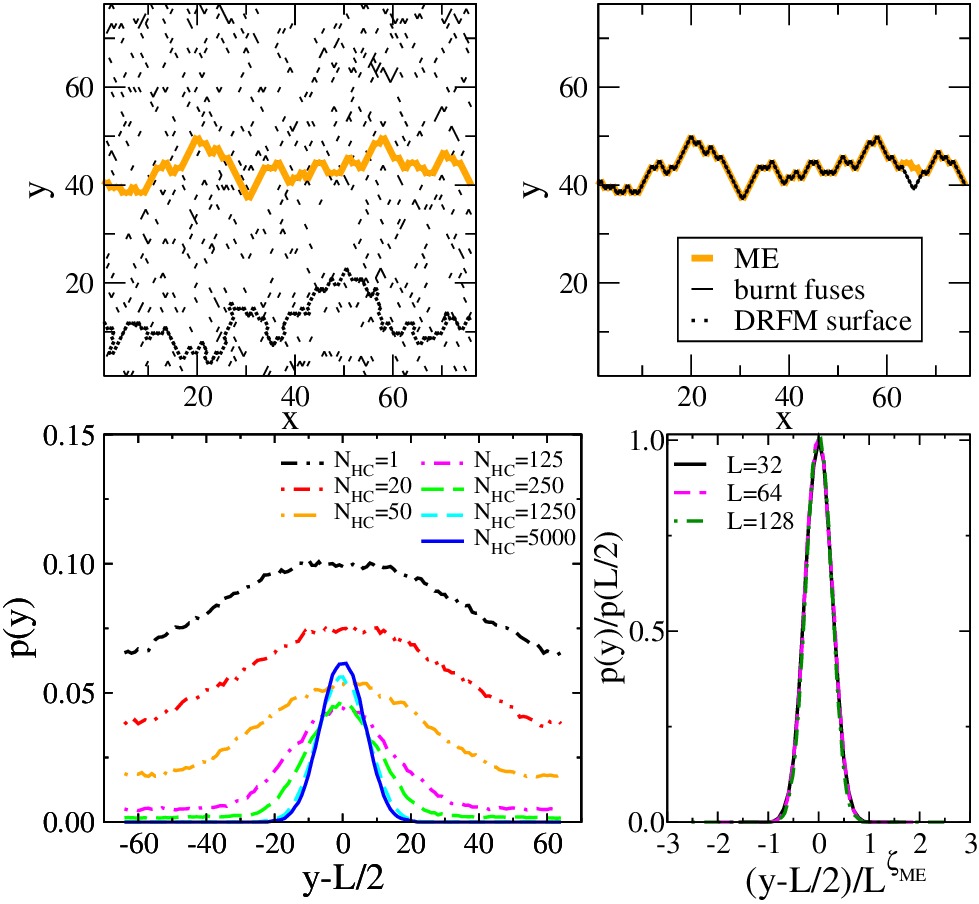}}
\caption{(Color online) Top: DRFM fracture surface in a system
of size $L=75$ for ${\cal N}_{HC}=1$ (left) and ${\cal N}_{HC}=10^{3}$ (right)
compared with the corresponding ME path for the same disorder realization.
Bottom panels show the average damage profiles for the DRFM in a system of size
$L=128$ for different values of ${\cal N}_{HC}$ (left) and the corresponding
data collapse for different system sizes in the case of large ductility ${\cal
N}_{HC}=5 \times 10^3$ (right).}
\label{fig3}
\end{figure}

\paragraph{Strain localization.-} The dynamics of damage
accumulation in the DRFM is summarized in Fig.~\ref{fig3}. For
quasi-brittle conditions, {\it i.e.} when we induce fracture after
only a few healing cycles (${\cal N}_{HC}=1$ in the left top panel
of Fig.~\ref{fig3}), the damage in the form of broken bonds is
spatially distributed throughout the sample in a randomly uniform
fashion, in analogy to the brittle RFM. In contrast, a ductile
sample can accumulate strain before any particular bond is broken.
Right top panel in Fig.~\ref{fig3} shows how damage strongly
localizes at the crack surface in a very ductile case with ${\cal
N}_{HC}=10^{3}$ healing cycles, which corresponds to an average
accumulated plastic strain of order $\epsilon_i \sim {\cal N}_{HC}
\beta T_i/(g_{i}L_y)$ per site. In this case, the crack strongly
localizes very close to the ME surface. Eventually, the damage
localizes exactly at the ME path as ${\cal N}_{HC} \to \infty$ (for
a finite but large ${\cal N}_{HC}$ in a finite sample,
cf.~Fig.~\ref{fig2})). In the DRFM the yield localization behaves
originally randomly except that, due to stress enhancements, a
degree of local ``clustering'' exists, and it increases slowly along
the stress-strain curve until the final localization of damage and
yielding as the fracture surface starts to be formed. Such a trend
is analogous to what is seen in the brittle RFM for damage
accumulation~\cite{alava06}. This is interesting since such local
plastic strains could be measured in experiments.

In the bottom panels of Fig.~\ref{fig3}, we depict the spatial distributions of
damage. The average damage profile $p(y)$ is calculated from the fraction of
broken bonds $n_b͑(y)$ along the $y$ direction and is computed as $p͑(y) =
\langle n_b͑(y)\rangle /L_x$, where the averaging is obtained by first shifting
the damage profiles by the center of mass and then averaging over different
samples \cite{nukala04}. The damage becomes narrower for larger ${\cal N}_{HC}$,
as the material is allowed to accumulate more local irreversible strain. Prior
to the growth of the final crack, there appears to be only local correlations in
the damage similarly in the brittle RFM \cite{nukala04,reurings05}, and the
maxima in the profiles arise from the crack path. The bottom right panel of
Fig.~\ref{fig3} shows the data collapse, $p(y) = p(L/2)
f[(y-L/2)/L^{\zeta_\mathrm{ME}}]$, for the average profiles of accumulated
damage for very ductile samples (${\cal N}_{HC} = 5 \times 10^3$) with the local
roughness exponent $\zeta_\mathrm{ME} =2/3$ corresponding to ME. This quantifies
how with an increased number of healing cycles fractures tend to only occur in
the final crack surface, which progressively tends to the ME path. We also
observe (not shown) that in the quasi-brittle limit damage is
volume-like, as expected for RFM~\cite{nukala04}, while in the extremely ductile
case damage scales as $\sim L$, as the failing elements are located
at the one-dimensional crack.
\begin{figure}
\centerline{\includegraphics[width=75mm,type=eps,ext=.eps,read=.eps]{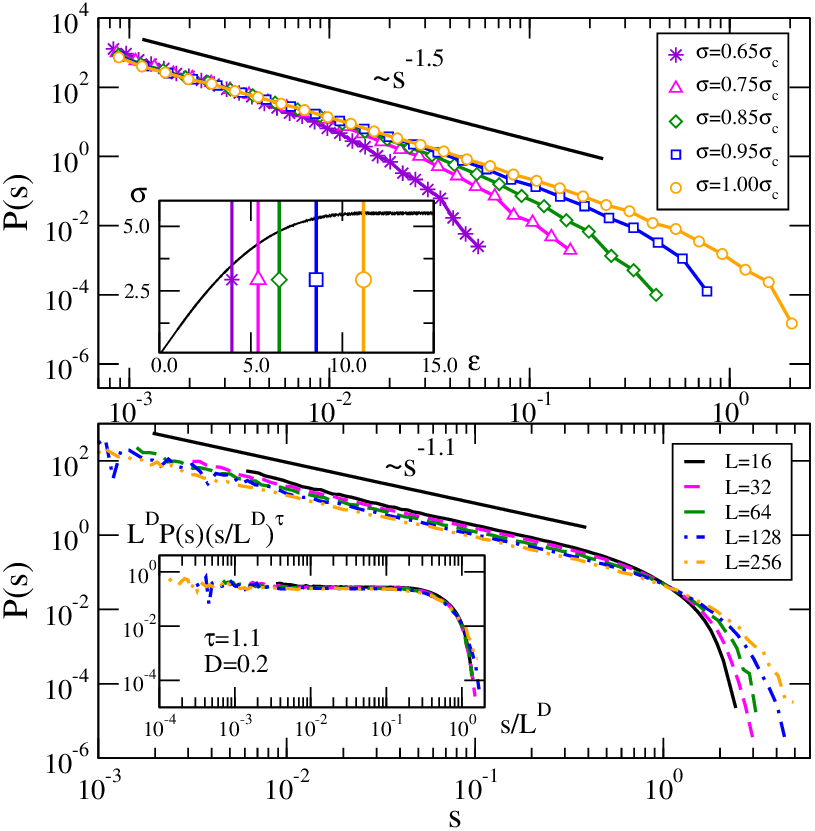}}
\caption{(Color online) Statistics of strain avalanches in the DRFM for
$\beta=0.1$ in a very ductile sample. Top: Avalanche distribution for a single
realization in a system of size $L=128$ collected up to different times in the
evolution as marked in the inset. Bottom: Avalanche distribution in the plastic
steady-state, well above the yielding point, for different system sizes averaged
over realizations of disorder. Inset shows the corresponding data collapse.}
\label{fig4}
\end{figure}

\paragraph{Avalanches of plastic events.-}
Indirect evidence by means of acoustic emission
experiments~\cite{miguel01, richeton06,wang09}, and very recent
direct micro-compression
measurements~\cite{dimiduk06,schwer07,zaiser08}, have shown that
strain bursts take place in the form of avalanches whose size
distribution decays as a power-law with an exponent that lies within
the range $\tau=1.4-1.6$. On the other hand, theoretical
arguments~\cite{zaiser05} and simulations in dislocation
models~\cite{miguel01,zap07,miguel08} and amorphous
materials~\cite{sor93b} suggest a universal exponent $\tau=3/2$. We
measured strain avalanches as the total strain occurred in the system
between two external stress increments. In the DRFM the avalanche
size is $s=\sum_{i=1}^{n}\Delta_{i}/L_y$, which corresponds to the
sum of the voltage sources $\Delta_{i}$ added between two external
voltage increments and $n$ is the number of fuses involved in the
avalanche. Figure~\ref{fig4} (top panel) shows how the distribution
of strain bursts for a sample of size $L=128$ evolves towards the
yielding point. As it can be observed the cutoff increases as the
stress reaches the critical point, with an exponent that approaches
$\tau = 3/2$. This is excellent agreement with recent measurements 
of the distribution of shear avalanches in ductile 
metallic glasses~\cite{sun10}.
On the other hand, well above the yielding point the
material response is fully plastic and we measure a different strain
avalanche distribution in this plastic steady-state, {\it i.e.} when
the strain has already become completely localized and the average
response in the global stress-strain curve in Fig.~\ref{fig1} is
constant. To do this we start to record statistics of strain
avalanches well above the yielding point and obtain $\mathcal{P} (s)
= L^{-D}(s/L^D)^{-\tau}f(s/L^{D})$ with $\tau=1.1 \pm 0.01$ and
$D=0.2\pm 0.01$, where $f$ is a finite-size scaling function
obtained by a data collapse for different system sizes (bottom
panel, Fig.~\ref{fig4}). $\tau \to 1$ corresponds to the plastic
flow regime where the path of yielding sites effectively separates
the system into two parts.

In summary, we have introduced a lattice model of elasto-plastic disordered
materials which yield and finally fracture. The model exhibits a transition from
purely brittle to fully ductile fracture, and includes both limits. A small
accumulation of yield strain corresponds to quasi-brittle behavior. In contrast,
as ductility is increased the resulting fracture paths gradually approach ME
surfaces, and the damage decreases. The yielding process takes place in the form
of avalanches of strain events that are found to be power-law distributed with
an exponent $\tau \approx 1.5$, which is to be compared with experiments
reporting
$\tau=1.4-1.6$~\cite{miguel01,richeton06,wang09,dimiduk06,schwer07,zaiser08,
sun10} and close to an earlier theoretical prediction, $\tau =
3/2$~\cite{zaiser05}. Interestingly, the DRFM also allows us to investigate
avalanche dynamics in the plastic steady-state regime [$\sigma \sim
O(\epsilon^0)]$ where we found $\tau \approx 1.1$, consistent with plastic flow
and unbounded strain events in response to the external loading. Future
directions from here would include studying the DRFM in three dimensions, and
adding strain-hardening to the non-linearly elastic part of the element
response.

\acknowledgments
We thank financial support from
MICINN (Spain) through project No. FIS2009-12964-C05-05,
the FPU program of MEC (Spain),
ISI foundation (Italy), the Center of
Excellence program of the Academy of Finland, and the European Commission
through contract NEST-2005-PATH-COM-043386.


\end{document}